\documentclass[aps, twocolumn, letterpaper]{revtex4}
\usepackage{natbib}
\usepackage{graphicx}
\bibpunct{}{}{,}{s}{}{\textsuperscript{,}} 
\newenvironment{frontmatter}{}{\maketitle}
\newcommand{\ead}[1]{}
\newcommand{\corauthref}[1]{}
\newcommand{\corauth}[2][x]{\thanks{#2}}
\newenvironment{keyword}{Key words: }{}
\newcommand{\sep}{; }
\renewcommand{\address}[1]{\affiliation{#1}}

\begin{document}
\begin{frontmatter}

\title{Protein Secondary Structure: Entropy, Correlations and Prediction}

\author{Gavin E. Crooks\corauthref{gec}}
\ead{gec@compbio.berkeley.edu}
\corauth[gec]{Corresponding Author:
	Gavin E. Crooks, 
	Department of Plant and Microbial Biology,
	111 Koshland Hall \#3102,
	University of California,
	Berkeley, CA 94720-3102, USA.
	Tel: +1 510-642-9614,
	Fax: +1 208-279-8978}
\author{Steven E. Brenner}
\address{Department of Plant and Microbial Biology,
University of California, Berkeley, CA, USA 94720-3102}

\begin{abstract}
Is protein secondary structure primarily determined by local interactions between residues closely spaced along the amino acid backbone, or by non-local tertiary interactions? 
To answer this question we have measured the entropy densities of primary structure and secondary structure sequences, and  the local inter-sequence mutual information density.
We find that the important inter-sequence interactions are short ranged, that correlations between neighboring amino acids are essentially uninformative,  and that only $1/4$  of the total information needed to determine the secondary structure is available from local inter-sequence correlations. 
Since the remaining information must come from non-local interactions, this observation supports the view that the majority of most proteins fold via a cooperative process where secondary and tertiary structure form concurrently.
%
To provide a more direct comparison to existing secondary structure prediction methods, we construct a simple hidden Markov model (HMM) of the sequences. This HMM achieves a prediction accuracy comparable to other single sequence secondary structure prediction algorithms, and can extract  almost all of the inter-sequence mutual information.
This suggests that these algorithms are almost optimal, and that we should not expect a dramatic improvement in prediction accuracy. 
%
However, local correlations between secondary and primary structure are probably of 
under-appreciated importance in many tertiary structure prediction methods, such as threading.
%

\end{abstract}

\end{frontmatter}

\section*{INTRODUCTION}

The secondary structure of a protein is a summary of the general conformation and hydrogen bonding pattern of the amino acid backbone\cite{FrishmanArgos-1995-Proteins-STRIDE}. This structure is frequently simplified to a sequence (one element per residue) of helixes (H), extended strands (E) and unstructured loops (L).
It has long been recognized that each residue's secondary structure is appreciably correlated with the local amino acid sequence\cite{Szent-Gyorgyi-1957-Science} and that these correlations may be used to predict the secondary structure\cite{Rost-2001-JStructBiol,PrzybylskiRost-2002-Proteins}, or as a contribution to threading potentials\cite{Alexandrov-1996,McGuffin-2003-Bioinformatics} and other tertiary structure prediction algorithms\cite{Bowie-1991-Science}.
The effectiveness of local secondary structure prediction,  and the utility of secondary structure potentials, depends upon the extent 
to which a protein's structure, particularly the secondary structure, is determined by local, 
short-ranged interactions between residues closely spaced along the backbone, as opposed to non-local or long-ranged tertiary interactions.

\begin{figure}[t]
\includegraphics{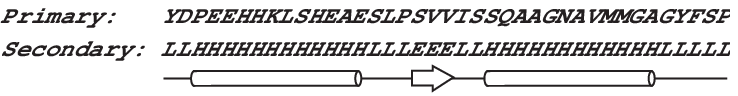}
\caption{A protein's amino acid sequence is correlated with the corresponding secondary structure sequence, represented here by a sequence of helixes (H), extended strands (E) and unstructured loops (L). For example, alanines (A) are typically associated with helixes, while glycines (G) are often located near helix breaks. Also note that secondary structure is strongly persistent. Helixes, for example, are on average about 10 residues long\cite{Schmidler-2000-JCompBio}.  
}
\label{seq}
\end{figure}

The strength, organization and relative importance of local sequence-structure interactions can be determined with a statistical analysis of the corpus of known protein structures. 
We treat the primary and secondary structures of a protein as random 
sequences composed from either the 20 letter amino acid or the 3 letter EHL (Extended strand / Helix / Other) structure alphabets, as shown in Fig.~\ref{seq}.  These sequences contain substantial local sequence and inter-sequence correlations which can be quantified using entropic measures\cite{CoverThomas}. To ensure accurate results we employ a large, carefully curated collection of protein structures derived from the
Structural Classification Of Proteins (SCOP)\cite{Murzin-1995-JMB,Lo_Conte-2002-NAR} database, that contains 2,853 sequences.

\subsection*{Sequence Information}

Entropy is  a measure of the information needed to describe a random variable\cite{CoverThomas}. Specifically, 
the entropy $H(X)$ of a discrete random variable $X$, measured in bits, is defined as
\begin{equation}
    H(X) = - \mathrm{E}\big( \log_2 P(X) \big) = -\sum_{x\in{\mathcal X} } P(x) \log_2 P(x) ,
\end{equation}
\noindent where $\mathcal X$ is the alphabet, the set of allowed states, $x$ is an element of 
$\mathcal X$, $\mathrm{E}(X)$ is the expectation,  and $P(x)$ is the probability of state $x$. 
When considering the  entropy of a collection of variables it is important to take into account inter-variable correlations.  For a statistically homogeneous random sequence with local correlations the appropriate information measure is the entropy density $h_\mu$, the rate at which the entropy of the sequence increases with length:
\begin{equation}
   h_{\mu} =  \lim_{L\rightarrow\infty} \frac{H(X^L) -E_h}{L} 
\label{EntropyDensity}
\end{equation}
\noindent Here, $H(X^L)$ is the entropy of sequence fragments, $X^L$, of length $L$. The non-extensive excess entropy, $E_h$, is the quantity of information explained away by taking account of inter-site correlations.  The entropy density is also referred to as the entropy rate or metric entropy\cite{CoverThomas}.

 A convenient measure of correlation between two discrete random variables, $X$ and $Y$, is the  
mutual information $I(X;Y)$, defined as 
\begin{eqnarray}
   I(X;Y) &=& H(X) + H(Y) - H(X,Y) ,\\
&=& \sum_{x\in{\mathcal X}, y\in{\mathcal Y} } P(x,y) \log_2 \frac{P(x,y)}{P(x)P(y)} ,
\end{eqnarray}
\noindent where $P(x,y)$ is the joint probability of observing states $x$ and $y$. 
 If the random variables are independent ($P(x,y)=P(x)P(y)$) then the mutual information achieves its lower bound of zero.  Mutual information cannot exceed the entropy of either variable, and this upper bound is reached when the variables are perfectly correlated ($P(x,y)=P(x)=P(y)$). 

The appropriate entropic correlation measure for a pair of statistically homogeneous random sequences is the mutual information density, $i_{\mu}$, 
\begin{equation}
   i_{\mu} =  \lim_{L\rightarrow\infty} \frac{I(X^L;Y^L) -E_i}{L} .
\label{MutualInfoDensity}
\end{equation}
\noindent Here, $E_i$ is the excess mutual information.

When we consider the correlations between three random variables, it is often useful to consider $I(X;Y|Z)$, the conditional mutual information\cite{CoverThomas}  of $X$ and $Y$, given a third variable,  $Z$. This quantity can be conveniently defined in terms of mutual information
\begin{equation}
I( X; Y | Z) =  I(X;Y) + I( X,Y ; Z) - I(X;Z) - I(Y;Z) 
\label{ConditionalMutualInformation}
\end{equation}
Conditioning on a third random variable may increase or decrease the mutual information\cite{CoverThomas}.

\begin{table}[pt]
\caption{Summary of Primary Structure ($R$) and Secondary Structure ($S$) Sequence and Inter-Sequence Information Measures. }
\centering
\begin{tabular}{lllll}
\\
PRIMARY && bits &&\\
residue entropy & $H(R_i)$       & $4.179$&$ \pm0.001$  &    \\
neighbor mutual info.& $I(R_i; R_{i+1})$& $0.006$&$\pm0.002$            &                   \\
conditional neighbor MI & $I(R_i; R_{i+1} | S_i S_{i+1})$ & $0.0159$&$\pm0.0004$ &\\ 
entropy density & $h_{\mu}(R)$ & $4.173$&$\pm0.003$&\\
\\
SECONDARY &&&&\\
residue entropy &$H(S_i)$       & $1.533$&$\pm0.002$ & \\
neighbor mutual info. & $I(S_i;S_{i+1})$ & $0.893$&$\pm0.003$                  &\\ 
entropy density &$h_{\mu}(S)$ & $0.598$&$\pm0.001$&\\
excess entropy &$E_h(S)$    & $0.997$&$\pm0.005$ &\\
\\
INTER-SEQUENCE \\
monomer mutual info.  & $I(R_i;S_i)$ & $0.0813$&$\pm0.0007$ &\\
dipeptide mutual info. & $I(R_i R_{i+1};S_i S_{i+1})$ & $0.208$&$\pm0.002$ &   \\
mutual info. density & $i_{\mu}(R;S)$ & $0.164$&$\pm0.003$&\\
\end{tabular}
\label{entropies}
\end{table}

\begin{figure}[t]
\centerline{
\includegraphics{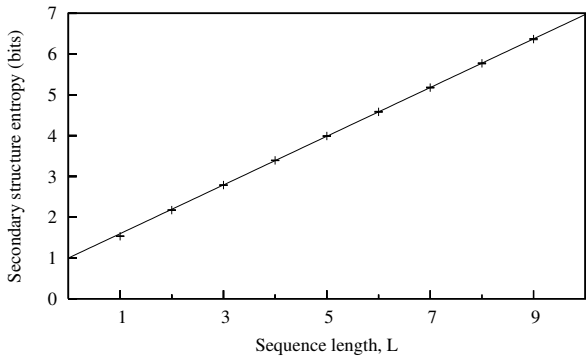}
}
\caption{Secondary structure sequences are strongly correlated, but the correlations have a simple structure. In this figure we plot the entropy of secondary structure blocks, $H(S^L)$,  as a function of block length, $L$ (points). Bootstrapped confidence intervals are smaller than the data point symbols. The linear increase of block entropies is indicative of a simple sequence, one that can, to a good first order approximation, be modeled as a low order Markov chain. A linear regression to the data (solid line) gives an excess entropy of $E_h \approx 1.0$ bits (zero intercept) and a true secondary structure entropy density of $h_\mu\approx 0.60$ bits per residue.
Over half of the single site entropy is explained away when we look beyond single site statistics.
}
\label{BlockEntropies}
\end{figure}

\begin{figure}[t]
\centerline{
\includegraphics{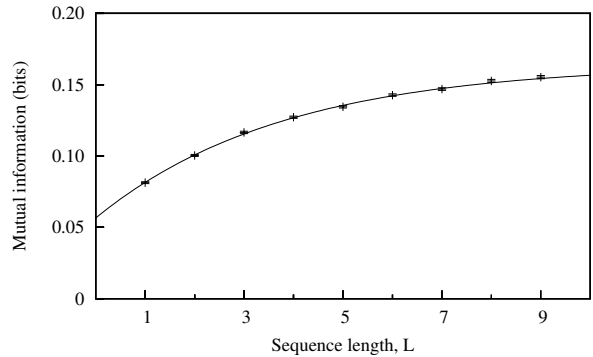}
}
\caption{The direct local interactions between primary and secondary structure are short ranged. Here, the mutual information, $I_c(R^1, S^L)$, between a block of secondary structure of length $L$ and the single amino acid located at the center (odd $L$) or immediately left of center (even $L$) of that block is plotted against block length (points). Bootstrapped confidence intervals are smaller than the data point symbols. A non-linear regression to an empirical exponential functional form gives a characteristic length scale of about 4 residues, and a limiting value of  $I_c(R^1,S^{\infty})  \approx 0.164$, which is a reasonable approximation to the total mutual information density, $i_{\mu}(R;S)$.
}
\label{MI}
\end{figure}

\section*{RESULTS}

\subsection*{Entropy and Correlations}

In Fig.~\ref{BlockEntropies} we plot the entropies for secondary structure sequence blocks up to length 9 ($3^{9}= 19683$ states).
Of the half million residues in our data set, about 23\% are assigned to strand, 39\% to helix, and 38\% to other, resulting in a relatively large single site secondary structure entropy of 1.53 bits. (The maximum entropy for three states is  $\log_2 3\approx 1.59$ bits.) However, neighboring secondary structure elements are strongly correlated, resulting in a relatively large nearest neighbor mutual information, $I(S_i;S_{i+1})\approx0.89$ bits. A linear regression to the asymptotic functional form, $H(S^L) \sim L h_{\mu} + E_h$ ($L\geq3$) gives an excess entropy of $E_h =0.997\pm0.004$ bits, and an entropy density of 
$h_{\mu}= 0.598\pm0.001$ bits per residue. This entropy density, the amount of information needed to describe the secondary structure sequence,  is considerable less than the single site entropy ($1.53$ bits) due to the strong inter-site correlations that may be observed in Fig.~\ref{seq}. 

It is notable that the entropies for short blocks are almost identical to  the 
asymptotic linear extrapolation used to estimate entropy density and excess entropy 
(Fig.~\ref{BlockEntropies}).  This property is indicative of a sequence with a simple structure, and suggests that many of the important statistical features of secondary structure sequences can be  successfully modeled by a low order Markov chain\cite{CrutchfieldFeldman-2003}.

In contrast to secondary structure, neighboring amino acids are only weakly correlated. The nearest neighbor mutual information, $I(R_i;R_{i+1})\approx0.006$ bits, is small relative to the single site entropy  of $H(R_i)\approx4.18$ bits,
which, consequentially,  is almost identical to the primary sequence entropy density.  Moreover, the mutual information between neighboring amino acids, conditioned upon the  corresponding secondary structure (Eq.~\ref{ConditionalMutualInformation}), is also relatively insignificant: $I(R_i;R_{i+1}|S_i S_{i+1} )\approx0.016$ bits.  
Neighboring amino acids are approximately independent\cite{Weiss-2000-JTheorBiol}, irrespective of the local structure. The correlations between more distantly separated residues are also very small.

The strength of the primary to secondary structure sequence correlations
is quantified by the inter-sequence mutual information density.  However, the mutual information can only be directly calculated for
short sequence blocks due to the large effective alphabet of 60 ($=3\times20$) symbols. 
The observed single site mutual information is $I(S_i;R_i)\approx0.081$ bits, and the dipeptide mutual information is $I(R_{i}R_{i+1};S_{i}S_{i+1})\approx0.208$ bits, or $0.104$ bits per residue. 
Fortunately, to a good approximation we can neglect the correlations between amino acids, since neighboring residues are almost (conditionally) independent.   
For example, the dipeptide mutual information, $I(R_{i}R_{i+1};S_{i}S_{i+1})\approx0.208$ bits,  can be approximated
by $I(R_{i};S_{i}S_{i+1}) + I(R_{i+1};S_{i}S_{i+1})\approx0.198$, an expression that explicitly ignores amino acid correlations. The relatively small error of 0.010 bits (less than 5\% of the dipeptide  mutual  information) is directly related to the mutual information between neighboring amino acids, since (by Eq.~\ref{ConditionalMutualInformation})
\begin{eqnarray*}
I(R_{i}R_{i+1};S_{i}S_{i+1})-I(R_{i};S_{i}S_{i+1}) - I(R_{i+1};S_{i}S_{i+1})  
=\\
I(R_{i};R_{i+1}|S_{i}S_{i+1}) - I(R_i ; R_{i+1}) .
\end{eqnarray*}
\noindent It follows that the inter-sequence mutual information density can be estimated by examining $I_c(R^1; S^L)$, the mutual information between a block of secondary structure and the single amino acid located at the center of that block. (See Fig.~\ref{MI}.)
Empirically, we expected these entropies to decay exponentially towards their limiting value as block lengths increase\cite{CrutchfieldFeldman-2003}. A nonlinear regression to the functional form $a - b \exp(-L/c)$, (using data from odd block lengths only), gives $c=3.8\pm0.3$ residues for the characteristic length scale, $b = 0.108\pm0.002$ for the scaling prefactor  and $a=0.164\pm0.003$ bits for the central amino acid to secondary structure mutual information in the infinity block length limit. This last value is a good approximation to the inter-sequence mutual information density, $i_{\mu}(R;S)$,  with a bias, due to  neglecting amino acid correlations,  that is probably less than 10\%.

In summary, the direct local interactions are short ranged, neighboring amino acids are almost
independent, secondary structure sequences are correlated, but essentially Markovian, and the important inter-sequence correlations are local, with a characteristic length scale of about 4. The inherent information content of secondary structure sequences is $0.60$ bits per residue, about 4 times greater than the $0.16$ bits per residue of local mutual information between primary and secondary structure. These measurements place severe constraints on any single-sequence prediction algorithm that purports to extract secondary structure information from local sequence correlations. In particular, no analysis can extract additional information from the signal (the data processing inequality\cite{CoverThomas}) and  therefore, any sequence local prediction of secondary structure can contain no more information than that contained in the local primary-secondary sequence correlations.

\subsection*{Prediction}

\begin{figure}[t]
\centerline{
\includegraphics{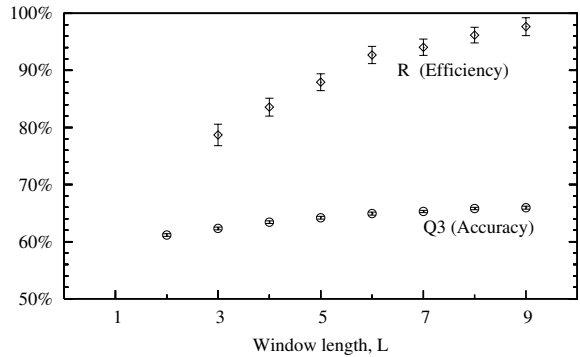}
}
\caption{
The hidden Markov model defined in  Eqs.~\ref{bayes}-\ref{markov} is able to extract over 95\% of the available inter-sequence information. Here, the efficiency, $R=i_{\mu}^{\rm HMM}/i_{\mu}$ and average 3 state accuracy ($Q_3$) are plotted against HMM window size ($L= k+1$)  for single sequence prediction on the SCOP 1.61 40\% STRIDE/CK data set.  Results for the Barton data set are similar. Window sizes cannot be reliably extended beyond those shown here due to finite sequence data.  The model information density, $i_{\mu}^{\rm HMM}$, approaches (but cannot exceed) the inter-sequence mutual information density, $i_{\mu}$, indicating that the model is almost optimal. The prediction accuracy  $Q_3 =65.9\pm0.3\%$ at $L=9$, is the same (within statistical errors) as the accuracy of a variety of comparable secondary structure prediction algorithms, suggesting that these algorithms are also almost optimal.
}
\label{predR}
\end{figure}

Many different algorithms have been proposed for predicting secondary structure from local inter-sequence correlations. Interestingly, the underlying organization of the majority of these algorithms does not reflect the underlying organization of the intra-  and inter-sequence interactions elucidated in the preceding section. Typically, these methods use a large primary structure window of around 15 to 27 residues to predict the single secondary structure element at the center of that window, and often assume that inter-amino acid correlations are informative. However,  even nearest neighboring amino acids on the chain are only weakly correlated, and these correlations provide negligible information about the local structure.

As an alternative prediction algorithm, we have constructued a relatively simple hidden Markov model (HMM) (Eqs.~\ref{bayes}-\ref{markov}, Fig.~\ref{FactorGraph}) that embodies three key approximations;  that protein sequences are statistically homogeneous,  that direct secondary structure to primary structure interactions are local along the chain, and that amino acids at neighboring sites are independent. Instead of a large primary structure window, we use short, overlapping secondary structure windows.  Similar models, with similar assumptions,  can be found in the work of Thompson and Goldstein\cite{ThompsonGoldstein-ProtienSci-1997} and Schmidler {\it et al}.\cite{Schmidler-2000-JCompBio}.

We estimate the amount of information that the HMM successfully extracts by measuring the mean log odds of the observed secondary structure fragments (Eq.~\ref{logodds}), and then extrapolating across different length scales to estimate the model mutual information density, $i_{\mu}^{\mathrm{HMM}}$ (Eq.~\ref{MutualInfoDensity}). 
Since the maximum amount of information that can be extracted is the previously estimated inter-sequence mutual information density $i_{\mu}(R;S)$ (Fig.~\ref{MI}),  we may profitably consider the efficiency ratio,  $R=i_{\mu}^{\rm HMM}/i_{\mu}$, which is plotted in figure~\ref{predR}. This model is able to extract over $90\%$ of the available information with a modest secondary structure window size of only $L=7$. In other words, the prediction algorithm is almost optimal. 

The most common measure of secondary structure prediction quality is the average three state accuracy, $Q_3$, the average fraction of residues that are correctly classified as helix, strand or other. Prediction accuracy increases monotonically with window length, reaching $65.9\pm0.3\%$ at $L=9$ (See Fig.~\ref{predR}). We cannot reliably increase the window size further due to the finite size of the training and test data sets.

Prediction accuracy can vary considerably due to variations in secondary structure assignment and due to variations in the underlying data set itself. Our standard data set  consists of 2,853 sequences derived from the 40\% subset of SCOP release 1.61, with STRIDE\cite{FrishmanArgos-1995-Proteins-STRIDE} secondary structure assignments.   We also considered prediction accuracies for the Cuff-Barton\cite{CuffBarton-2000-Proteins} library of 513 sequences, using STRIDE and DSSP\cite{KabschSander-1983-Biopolymers-DSSP} secondary structure assignments, and two different reductions of the STRIDE and DSSP alphabets to 3 states, the CK and EHL mappings (For details, see Materials and Methods).  At $L=7$ accuracy ranges from $63.6\pm0.6\%$ to $66.4\pm0.7\%$. The maximum accuracy is achieved with the CK mapping, irrespective of the secondary structure assignment program. Essentially, the CK mapping produces more coherent, less random secondary structure sequences than the EHL mapping, which leads to more facile prediction.
Using the smaller library of only 513 sequences leads to substantial standard errors of about $0.6$\%, and to a large estimated bias of about $0.7$\%. (Without a bias correction our maximum reported accuracy would be $67\%$.) A number of different secondary structure prediction algorithms have been tested upon the Barton data set. However, given these small sample errors and the variation due to changes in secondary structure assignment, we cannot statistically distinguish accuracies separated by less than about 2 points\cite{Rost-2001-Proteins-EVA,PrzybylskiRost-2002-Proteins}. Since the range of reported accuracies is about 65\%-68\%\cite{CuffBarton-1999-Proteins,Schmidler-2000-JCompBio,Kloczkowski-2002-Proteins-GORV}, we are obliged to conclude that many, very different secondary structure prediction algorithms are statistically indistinguishable.

Our HMM model is almost optimal, in the sense that it extracts almost all of the available information.  Moreover, the accuracy of our model is approximately the same (within statistical and systematic errors) as the maximum accuracy of a variety of other secondary structure prediction methods that utilize only local sequence-sequence correlations\cite{Chandonia-1996-ProtSci,CuffBarton-1999-Proteins,Schmidler-2000-JCompBio,Kloczkowski-2002-Proteins-GORV}. This suggests that these algorithms are also almost optimal, and that the modest prediction accuracy is due to the fundamental lack of local structure information. Conversely, the fact that these diverse, sophisticated prediction algorithms are not able to extract additional signal from local correlations indicates that we have not overlooked some subtle source of secondary structure information in our analysis of local inter-sequence correlations.

It has been found that 
secondary structure prediction accuracy can be substantially enhanced by basing the prediction upon a multiple sequence alignment (MSA) of homologous protein sequences\cite{RostSander-1993-JMolBiol}, rather than just a single sequence.  Since protein structure tends to evolve relatively slowly, the MSA essentially represents many, semi-independent amino acid sequences, each associated with approximately the same secondary structure sequence. 
How informative is this additional data? We extended our HMM to handle this evolutionary information (described in Eqs.~\ref{markov}-\ref{relative} ), and tested the model on the MSAs provided with the Barton data set\cite{CuffBarton-2000-Proteins}. This resulted in a three state accuracy of $72.2\pm0.6\%$, an improvement of about 6 points over the equivalent single sequence results. Since the information ratio for this data set was $R\approx1.3$,
 this modest accuracy increase actually represents a considerable increase in information. This accuracy is similar to reported accuracies of a number of other algorithms tested on this data set\cite{Schmidler-2000-JCompBio,Kloczkowski-2002-Proteins-GORV}. It may be that many profile based secondary structure prediction algorithms are essentially equivalent, and that the differing results are due, primarily, to differences in the quality of the input alignments\cite{Rost-2001-JStructBiol,Kloczkowski-2002-Proteins-GORV}.

\section*{DISCUSSION}

Although local inter-sequence information is insufficient to accurately determine secondary structure, such correlations are still useful to statistical tertiary structure prediction algorithms. For example, in protein threading\cite{Jones-1992-Nature-Threader} a primary sequence is matched to a structural template using an amino-acid contact potential, and other similar potentials derived from sequence-structure correlations.  Recently, the information contained in amino acid contacts was estimated to be about $0.04$ bits per contact, or $0.06$ bits per residue\cite{Cline-2002-Proteins}, which can be compared to our estimate of $0.16$ bits per residue of primary to secondary structure mutual information. Therefore, local structure potentials may be of under-appreciated importance to threading, and other similar statistical structure prediction methods.
Many such methods do consider secondary structure\cite{McGuffin-2003-Bioinformatics}, but some of these only consider the direct correlation between an amino acid and the secondary structure class at that one residue\cite{Bowie-1991-Science,Alexandrov-1996}. By ignoring the correlations between an amino acid and an extended segment of local secondary structure such methods lose over half of the available local signal, and,  unlike secondary structure prediction algorithms, are not optimal. 

Protein folding is also constrained by the scarcity of local structure information, since the mechanism by which information is extracted, either by a computer or physics, is irrelevant. Secondary structure must be predominately determined by non-local interactions, that in turn depend on the overall, native fold of the protein. But the native fold cannot be achieved until the native secondary structure has formed.
Therefore, protein folding must typically proceed by a cooperative mechanism\cite{Baldwin-1999-TIBS}, where secondary and tertiary structures form concurrently. Note, however, that since this conclusion is based upon a statistical analysis, it applies only to proteins on the average, and does not preclude particular proteins, or parts of proteins, from folding via a hierarchal mechanism\cite{Baldwin-1999-TIBS} where pre-organized local secondary structure elements collapse successively into ever-larger structures. 
For example, it has been suggested that the B-domain of {\it staphylococcal} protein A\cite{Myers-2001-NatStructBiol}, a small, single domain protein, can fold extremely quickly because of its strongly defined native secondary structure, which persists even in the unfolded state. If this is a general property of fast folding proteins, then the widely divergent folding rates of single domain proteins may be strongly correlated with the accuracy to which a particular proteins secondary structure can be predicted from the primary sequence.

There are at least two approaches to prediction that aim to circumvent the lack of local structure information. One is to utilize evolutionary information. Since protein sequences evolve more rapidly than protein structure, a multiple sequence alignment of a homologous family represents many, semi-independent sequence samples of approximately the same protein structure. Local structure prediction quality is then limited by the size of the family, the divergence of structure across the family, and the quality of the alignment. This strategy is commonly employed in secondary structure prediction\cite{RostSander-1993-JMolBiol}, and  improvements in accuracy to about $Q_3\approx75\%\pm3$ are routine\cite{CuffBarton-1999-Proteins,CuffBarton-2000-Proteins,ChandoniaKarplus-1999-Proteins,Rost-2001-JStructBiol,Eyrich-2003-Proteins-CASP5}.  By modifying our HMM to use evolutionary profiles, we find that even a modest increase in prediction accuracy represents a substantial increase in secondary structure information.

The alternative approach is to explicitly  incorporate non-local interactions. This is essentially what threading attempts to do, although the relatively small magnitude of contact potential information suggests that the bulk of non-local information is subtle, and difficult to extract. Of course, in principle we can determine the full three-dimensional structure of a protein using an atomic detailed molecular simulation. Until this becomes routinely feasible, computational structure determination will have to proceeded via less direct, statistical approaches.

\section*{MATERIALS AND METHODS}

\subsection*{Secondary Structure Library}

Ideally, a secondary structure library should be based upon a representative, high-quality and non-redundant subset of available protein structures. 
The Protein Data Bank  (PDB)\cite{Berman-2000-NAR-PDB} currently contains
contains over 20,000 publicly accessible structures, but many of these are very similar, and many are of relatively low quality.
The Structural Classification Of Proteins (SCOP)\cite{Murzin-1995-JMB,Lo_Conte-2002-NAR} database provides a convenient decomposition of PDB structures into domains, and the ASTRAL\cite{Brenner-2000-NAR-ASTRAL,Chandonia-2002-NAR-ASTRAL} compendium provides representative subsets of SCOP domains, filtered so that no two domains share more than a given percentage level of sequence identity. 
This filtering preferentially retains higher quality structures, as judged by AEROSPACI scores\cite{Chandonia-2002-NAR-ASTRAL}, an agglomeration of several structure quality measures.  
We selected the ASTRAL 40\% sequence identity subset of SCOP release 1.61, which was further filtered to remove multi-sequence domains, SCOP classes f (membrane and cell surface proteins) and g (small proteins), and retain only those structures determined by X-ray diffraction at better than 2.5~\AA\ resolution. 
The protein sequences were taken from the ASTRAL Rapid Access Format (RAF) sequence mappings\cite{Chandonia-2002-NAR-ASTRAL} which provides a more reliable and convenient representation of the true sequence than the PDB ATOM or SEQRES records. 
The secondary structure sequences were determined by the program STRIDE\cite{FrishmanArgos-1995-Proteins-STRIDE} using each protein's hydrogen bonding pattern and backbone torsional angles.
STRIDE was unable to process a small fraction of SCOP domains, which were consequentially removed from further consideration. The resulting library contains 2,853 protein domains and 553,373 residues.

For comparative purposes, we also studied the secondary structure library
of Cuff and Barton\cite{CuffBarton-1999-Proteins,CuffBarton-2000-Proteins}, which consists of 513 proteins and 84,091 residues. Secondary structure assignments are provided by both STRIDE and the program 
DSSP\cite{KabschSander-1983-Biopolymers-DSSP}. This data set also includes, for each structure, a multiple alignment of homologous sequences. These multiple sequence alignments were converted to amino acid probability profiles\cite{DurbinEddy-1998} using the program 
{\tt hmmbuild} from HMMER (v2.3)\cite{HMMER}.

Both DSSP and STRIDE assign each residue's secondary structure to one of 8 classes;
$\alpha$-helix (H), $3_{10}$ helix (G),  $\pi$-helix (I), $\beta$-strand (E),
$\beta$-bridge (B or b), Coil (C, L, or space), Turn (T) or  Bend (S).
Unstructured or poorly resolved regions of the protein are unassigned (X).
These 8 classes were reduced to the three letter alphabet, E (Extended strand), H (Helix), and L (Loop/Other) using the common CK
mapping\cite{ChandoniaKarplus-1995-ProteinSci,FrishmanArgos-1997-Proteins,CuffBarton-1999-Proteins} E$\rightarrow$E; H$\rightarrow$H;  all others$\rightarrow$L. 
We also considered another common reduction, the ``EHL'' mapping\cite{CASP4,Rost-2001-Proteins-EVA}
 E,~B$\rightarrow$E; H,~G, I$\rightarrow$H; all others$\rightarrow$L.


\subsection*{Entropy Estimation and Bias Correction}

The entropy of a discrete probability can be estimated by sampling from the distribution, and then replacing the true probabilities, $P(x)$, by the observed frequencies, $f(x) = n_x/N$. Here, $N$ is the total number of samples, and $n_x$ is the number of observations of  state $x$. 
A useful alternative approach is to construct an approximation of the true probabilities, $g(x)\approx P(x)$ (e.g. Eq.~\ref{markov}), and then estimate the entropy by the mean log likelihood of the data\cite{Moddemeijer-2000}.
\begin{eqnarray}
H(X) &=& \mathrm{E} \big( \log_2 P(X) \big) \nonumber \\
&\geq& \mathrm{E} \big(\log_2 g(X) \big) \approx \sum_{i=1}^N \frac{1}{N} \log_2 g(x_i)
\end{eqnarray}
\noindent Similarly, the mutual information can be related to the mean log odds, since
\begin{equation}
I(X;Y) = \mathrm{E} \left( \log_2 \frac{P(X,Y)}{P(X)P(Y)} \right) = \mathrm{E} \left( \log_2 \frac {P(X|Y)}{P(X)} \right).
\label{logodds}
\end{equation}

A serious problem with either approach is  that entropies estimated from limited amounts of data tend to be significantly biased\cite{Miller1955}, resulting in a systematic 
underestimation of the true entropy, or overestimation of the mutual information. 
We used non-parametric bootstrap resampling\cite{EfronTibshirani-1993-IntroToBootstrap}  to correct for this bias, and to estimate standard statistical errors. Fifty replicas of the original data are generated by  sampling, with replacement, from the available sequences.  
This resampling has associated systematic and random errors that are approximately the same as the errors introduced by the original finite sampling of sequences from the true random distribution.
These error estimates were not significantly improved when the number of replicas was increased from 50 to 500. The requisite pseudo-random numbers were drawn from the Mersenne Twister generator\cite{MersenneTwister,GSL}.


\subsection*{Secondary Structure Hidden Markov Model}

\begin{figure}[t]
\centerline{
\includegraphics{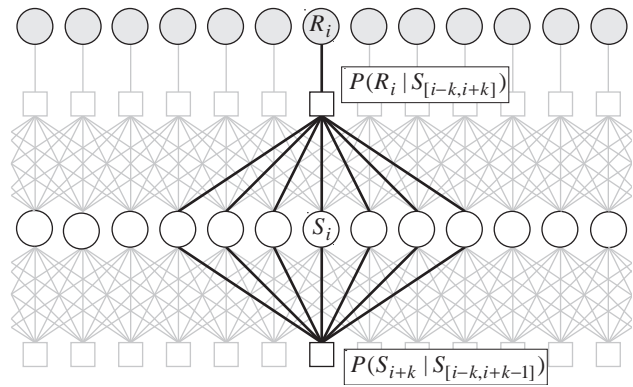}
}
\caption{A factor graph\cite{Kschischang-2001-IEEE} for $P(S|R)$, representing the decomposition of this complex,
many variable function into simpler parts (Eq.~\ref{bayes}-\ref{markov}). 
Circles represent variables  and squares represent factors, local functions of relatively few variables. The upper and lower rows of circles represent the primary and secondary structure sequences, respectively.
In this diagram $k = 2m=6$, for a window of size $k+1=7$. One set of factors, centered on sequence position $i$, have been highlighted.
The bottom factor connects $k+1$ neighboring secondary structure elements, and  represents the approximation of the 
secondary structure sequence probability by a $k$th order Markov chain (Eq.~\ref{markov}).
The factor between the chains represents the inter-sequence dependance (Eq.~\ref{amino}).
Thus, each residue is directly dependent upon a window of secondary structure (length $2m+1$), and is conditionally independent of neighboring residues.
}
\label{FactorGraph}
\end{figure}

The probability $P(S|R)$ of a secondary structure sequence, $S$, given the primary sequence, $R$, can be rewritten using Bayes' rule as 
\begin{equation}
P\left(S|R\right) = \frac{P\left(R|S\right) P\left(S\right)}{P\left(R\right)} .
\label{bayes}
\end{equation}
\noindent  Since the probability of an amino acid residue depends on the local secondary structure, and is almost independent of the identity of neighboring residues, to a good approximation the probability of each residue can be estimated from a short window of local secondary structure, 
\begin{equation}
P(R |S) \approx \prod_{i=1}^{L} P\left(R_{i} | S_{[i-m, i+m]}\right) .
\label{amino}
\end{equation}
\noindent Here, $X_i$ is the element at position $i$ and $X_{[i,j]}$ is a subsequence of length $i-j+1$ starting at position $i$ and ending at $j$.
Residues beyond the termini of the actual sequence ($i<1$, $i>L$)  are treated as undetermined.
The window size, $2m+1$, is an adjustable parameter of the model, and need not be particularly long, since the inter-sequence correlations have a characteristic length scale of only about 4 residues (See Fig.~\ref{MI}).
We approximate the prior probability of the secondary structure sequence by a $k$th order Markov chain,
\begin{equation}
P(S) \approx P\left(S_{[1,k]}\right)  \prod_{i=1}^{L-k} P\left(S_{i+k} | S_{[i,i+k-1]} \right).
\label{markov}
\end{equation}
\noindent The primary structure sequence probabilities $P(R)$ can be determined from normalization.  Combining the preceding approximations, (Eq.~\ref{amino} and \ref{markov}), using $k=2m$ for consistency, generates a hidden Markov model (summarized in Fig.~\ref{FactorGraph}) that emits the primary structure sequence on transitions between blocks of secondary structure of length $k$.

The probabilistic model of Eqs.~\ref{bayes}-\ref{markov} can be generalized so that the prediction is based upon a multiple sequence alignment (MSA) of homologous protein sequences. First, we convert the multiple sequence alignment into an amino acid profile, $\theta = \{\theta_1(r), \theta_2(r), \ldots, \theta_L(r)\},$
that represents the probabilities of each amino acid at each position of the protein of interest\cite{DurbinEddy-1998}.
The secondary structure probability, given this profile, may then be approximated as
\begin{eqnarray}
P\left(S|\theta\right) &=& \frac{P\left(\theta|S\right) P\left(S\right)}{P\left(\theta\right)}  ,\\
P(\theta |S) &\approx& \prod_{i=1}^{L} P\left(\theta_{i} | S_{[i-m, i+m]}\right) .
\end{eqnarray}

We expect that each residue's observed homology profile, $\theta_i(r)$, will vary from the structure profile, $P(r|S_{[i-m, i+m]})$, due to sampling errors, random site-to-site variation, inter-protein structural variation and because each residue is under different structural, functional and evolutionary constraints.
As a simple approximation, we use the large deviation distribution\cite{CoverThomas} to model the variation of the observed profile from the expected profile;
\begin{eqnarray}
\lefteqn{P(\theta_i | \beta,S_{[i-m, i+m]})}  \nonumber \\ &&\approx 
 \exp \Big\{- \beta D\big( \theta_i(r)  \big\| P(r|S_{[i-m, i+m]}) \big) \Big\} .
\label{relative}
\end{eqnarray}
\noindent Here, $D(p\|q)=\sum_i p_i \ln(p_i/q_i)$ is the relative entropy.
We treat $\beta$ as an empirical dispersion parameter that is independent of the secondary structure or primary structure profile. 

Computationally, the conditional secondary structure probabilities can be derived from the amino acid sequence using the standard forward-backward dynamic programming algorithm\cite{Rabiner-1989}.
The time and memory complexities for a naive implementation are $O(L 3^k)$, which, despite the exponential scaling, is feasible for moderate~$k$. For example, with $k=7$ training on one half of our library (2853 sequences) required 4 seconds from a modest contemporary PC (667 MHz PowerPC G4), and prediction of the other half required approximately 5 minutes, or about 5 sequences per second. 
In principle, a more efficient implementation is possible, since, although the total number of secondary structure sequences scales as $3^L$, the number of typical sequences with non-negligible probability scales as $2^{H(S^L)} \approx {1.5}^L$, by the asymptotic equipartition principle\cite{CoverThomas}. 
The optimal prediction at a particular site
is the secondary structure element with the greatest posterior probability.

The available sequence data was partitioned every other sequence into disjoint test and training sets of approximately equal size. The training set was used to estimate secondary structure block probabilities, $P\left(S_{[i, i+k]}\right)$ (regularized with a Laplace pseudocount of 1) 
and corresponding amino acid profiles,  $P\left(R_{i} | S_{[i-m, i+m]}\right)$ (regularized with a pseudocount of $20$ times  the amino acid background probability). 
Statistical errors were estimated from a full bootstrap resampling of both the test and training sequences.

\subsection*{Avaliability}
Both the data sets and {\tt second-hmm}, the program developed for this analysis, are freely available from our web site at 
{\tt http://compbio.berkeley.edu/}.

\section*{ACKNOWLEDGMENTS}
We would like to thank John-Marc Chandonia, Barbara Engelhardt, Mauro Merolle, Richard E. Green  and Avery A. Brooks for helpful discussions and suggestions, and Emma Hill and Jason Wolfe for critical readings of this manuscript.  Financial support was provided by the NIH (1-K22-HG00056) and the Sloan postdoctoral fellowship in computational molecular biology.

\bibliography{secondaryStr}

 \end{document}